\documentclass{cjpsuppl}    

\newcommand{\ci}[1]{\cite{#1}}

\newcommand{\baa}{\begin{eqnarray*}}

\newcommand{\eaa}{\end{eqnarray*}}

\newcommand{\eb}{\end{thebibliography}}

\begin{document}
\title{Sources of time reversal odd spin asymmetries in QCD }



\authori{O.V. Teryaev  
}       

\addressi{BLTPh, JINR, Dubna, Russia }

\authorii{}    \addressii{}
\authoriii{}   \addressiii{}
\authoriv{}    \addressiv{}
\authorv{}     \addressv{}
\authorvi{}    \addressvi{}

\headtitle{Sources of T-odd spin asymmetries  in QCD }

\headauthor{O.V. Teryaev }  
\lastevenhead{O.V. Teryaev: Sources of T-odd spin asymmetries in QCD }
\pacs{} 
\keywords{single spin asymmetry, distribution, fragmentation}

\refnum{}

\daterec{20 October 2002}    


\suppl{A}  \year{2003}

\setcounter{page}{1}


\maketitle

\begin{abstract}



Generation of T-odd single spin asymmetries (SSA) by the various ingredients 
of QCD factorization is discussed. The possible use of SSA in studies 
of Generalized Parton Distribution (GPD) at HERA with the polarized lepton beam is 
suggested. The role of GPD in the investigation of orbital angular momenta of partons 
is discussed. The generalization of Equivalence principle, leading to the equipartition 
of momenta and total angular momenta, violated in perturbation theory, 
but possibly restored due to confinement and chiral symmetry breaking, is proposed. 
The T-odd fragmentation and fracture function are considered. The T-odd distribution
(Sivers) function may be only effective, due to the imaginary cuts in the SIDIS and Drell-Yan process,
while the existence of such universal functions should lead, after the integration over transverse momentum,
to the strong T  violation in polarized DIS.



\end{abstract}

\section{Introduction}     



















The single transverse spin asymmetries are most easily studied
experimentally, as they require only one polarized particle
(usually, the polarized target). At the same time, they
are known to be one of the most subtle
effects in QCD. They should be proportional to mass scale,
and the only scale in "naive" perturbative QCD is that of the current
quark mass.

The additional suppression 
comes
from the fact, that single asymmetries are related to the antisymmetric
part of the density matrix. Due to its hermiticity, the imaginary
part of scattering amplitude is relevant. As a result, the spin-dependent
contribution to the hard scattering cross section starts at the one-loop
level only. More exactly, it is due to the interference of the one-loop
spin-flip amplitude and Born non-flip one.
At the same time, the Born graphs provide a leading approximation
to the spin-averaged cross section and the asymmetry is proportional to
$\alpha_s$.

These spin effects are T-odd but not related to the CP-violation. 
They are generated by the term in the cross-section 
proportional to the polarization of {\it one} of the  spin-$1/2$ particles. 
If parity is conserved, such term should contain Levi-Civita tensor
which changes sign when the direction of time axis is changed,
that's why the effect is called T-odd. Sometimes such a transformation
is called "naive" T-reflection. 
At the same time, the "full" time reflection T contains also the
interchange of the initial and final states, so if this later
operation is non-trivial (which is guaranteed by the existence of the 
non-trivial phases in the S-matrix elements), the T-odd effects may
emerge in the T-invariant theory. If, for some reason, phases cannot
emerge, the real T(CP) violation is required, and the role of phases 
is taken by the complex couplings. The appearance of phases
is, in turn, related through the generalized optical theorem to the
existence of the cuts in some kinematical variables. 
The analysis of such cuts in the different ingredients of
QCD factorization can be the main tool of the unraveling of T-odd
effect in QCD.    


\section{Generalized Parton Distributions}

Generalized Parton Distributions (GPD) encode the information about the hadron 
structure relevant for hard exclusive processes in the most complete way    
and has been
the subject of extensive theoretical investigations for a few years.
The 
Deeply Virtual Compton Scattering (DVCS) \cite{DVCS1,DVCS2} is the
probably the cleanest among these hard process.
The SSA currently plays a major role in its investigation\cite{SSADV}.
The imaginary phase emerges in the handbag parton subprocess, while the interference 
is that of DVCS and BH amplitudes. As the latter is substantially larger at JLAB
and HERMES kinematics, SSA provides the most sensitive test for GPD, as it depends 
on the ratio $r=|M_{DVCS}|/|M_{BH}|$ linearly, rather than quadratically.
At the same time, SSA may be also studied at HERA in the collider more as well,
provided the polarized lepton beam is used at the forthcoming stages of experiments (see \cite{Val}).  
As BH process contribution is much smaller, the BH-DVCS asymmetry  is of the order $1/r$, 
and the interference between various DVCS amplitudes becomes important. 
In particular, the twist-3 contributions to DVCS \cite{Ani1} and exclusive 
vector meson production \cite{AT02} may be studied in such a way. 

Another important relation of GPD to nucleon spin structure is provided by relations of 
their moments to the  
particular matrix elements of Belinfante
energy-momentum tensors, and in, turn, to the 
total angular momenta of partons,
\begin{eqnarray}
      \langle p'| T_{q,g}^{\mu\nu} |p\rangle
       &=& \bar u(p') \Big[A_{q,g}(\Delta^2)
       \gamma^{(\mu} p^{\nu)} +
   B_{q,g}(\Delta^2) P^{(\mu} i\sigma^{\nu)\alpha}\Delta_\alpha/2M ] u(p),
\label{def}
\end{eqnarray}
where $P^\mu=(p^\mu+{p^\mu}')/2$, $\Delta^\mu = {p^\mu}'-p^\mu$,
and $u(p)$ is the nucleon spinor.
We dropped here the irrelevant terms of higher order in $\Delta$,
as well as containing $g^{\mu \nu}$, which will be discussed later.
The parton momenta and total angular momenta are:
\begin{eqnarray}
      P_{q, g} =  A_{q,g}(0), \nonumber \\
      J_{q, g} = {1\over 2} \left[A_{q,g}(0) + B_{q,g}(0)\right] \ .
\end{eqnarray}
Taking into account the conservation of momentum
and angular momentum
\begin{eqnarray}
\sum_{i=q,G} \int_0^1 dx x H_i (x,\xi,Q^2)=A_q(0)+A_g(0)= 1 \label{m} \\
\sum_{i=q,G} \int_0^1 dx x (H+E)_i (x,\xi,Q^2)=A_q(0)+B_q(0)+A_g(0)+B_g(0)=1 \label{am},
\end{eqnarray}
where we recalled the relation of formfactors to the moments of GPD $H,E$,
one can see that the
difference between partition of the momentum and orbital angular momentum
entirely comes from "anomalous" formfactors
($B_q(0)=-B_g(0)=\sum_{i=q} \int_0^1 dx x E_i (x,\xi,Q^2)$).
It may look surprising that the smalness of such a contribution may be related \cite{OT99}
to some generalization \cite{OT99,krim}
of famous Equivalence principle (which, as it is known, was developed by Einstein
just in Prague).
Indeed, matrix element of energy momentum tensor determines the nucleon behaviour in the external 
gravity field.

Let us start with a more common case of the interaction
with electromagnetic field described by the matrix element
of electromagnetic current,
\begin{eqnarray}
  M=\langle P'| J_{q}^{\mu} |P\rangle A_\mu (q).
\label{A}
\end{eqnarray}
This matrix element at a zero momentum transfer is fixed by the
fact that the interaction is due to the {\it local} U(1) symmetry,
whose {\it global} counterpart produces the conserved charge
(and, of course depends on the normalization of eigenvectors
$\langle P|P'\rangle=(2\pi)^3 2E \delta(\vec P-\vec P')$).
\begin{eqnarray}
\langle P| J_{q}^{\mu} |P\rangle=2e_q P^\mu,
\label{e}
\end{eqnarray}
so that in the rest frame the interaction is completely defined by
the scalar potential:
\begin{eqnarray}
\label{0e}
  M_0=\langle P| J_{q}^{\mu} |P\rangle A_\mu = 2e_q M \phi(q)
\label{A1}
\end{eqnarray}
At the same time, the interaction with a weak classical gravitational
field is:
\begin{eqnarray}
M=\frac{1}{2}\sum_{q,G} \langle P'| T_{q,G}^{\mu \nu} |P\rangle h_{\mu \nu} (q),
\label{h}
\end{eqnarray}
where $h$ is a deviation of the metric tensor from its Minkowski value.
The relative
factor 1/2, which will play a crucial role, comes from the fact
that the variation of the action with respect to the metric
produces an energy-momentum tensor with the coefficient 1/2,
(while,say, the variation with respect to classical source $A^\mu$
produces the current without such a coefficient). It is this
coefficient that guarantee the correct value for the
Newtonian limit fixed by the {\it global} translational invariance
\begin{eqnarray}
\sum_{q,G} \langle P| T_{i}^{\mu \nu} |P\rangle=2 P^\mu P^\nu,
\label{e2}
\end{eqnarray}
which, together with the approximation for $h$
(with factor of 2
having the geometrical origin) \cite{LL2}
\begin{eqnarray}
\label{h2}
h_{00}=2\phi (x)
\end{eqnarray}
results in the rest frame expression:
\begin{eqnarray}
\label{0g}
M_0=\frac{1}{2}
 \sum_{q,G} \langle P| T_{i}^{\mu \nu} |P\rangle h_{\mu \nu}=
2 M \cdot M \phi(q),
\label{A2}
\end{eqnarray}

where we used the same notation for gravitational and scalar
electromagnetic potentials, and identified normalization factor
$2M$ in order to make the similarity between (\ref{0e}) and (\ref{0g})
obvious. One can see that the interaction with gravitational
field is described by the charge, equal to the particle mass,
which is just the equivalence principle.
It is appearing here as a low energy theorem, rather than postulate.
The similarity with
electromagnetic case allows to clarify the origin of such a theorem,
suggesting, that the interaction with gravity is due to the {\it local}
counterpart of {\it global} symmetry, although it may be proved
starting just from the Lorentz invariance of the soft graviton
approximation \cite{wein64}.

The situation with the terms linear in $\Delta$ is different
for electromagnetism and gravity. While such a term is
defined by the specific dynamics in the electromagnetic case,
producing the anomalous magnetic moment, the similar contribution
in the gravitational case is entirely fixed by the angular momentum
conservation (\ref{am}),
which
was known in the context of gravity for more than 35 years \cite{KO,BD,HD}
\footnote{The reason is that the structure of
Poincare group
is more reach than
that of $U(1)$ group.}.
It
means,
in terms of the gravitational interaction, that
{\bf Anomalous Gravitomagnetic Moment (AGM) of any particle
is identically equal to
zero}.

As soon as the formfactors in spin-1/2 case differ from the ones for
the matrix element of vector current $J^\mu$ by the common factor
$P^\nu$, one may define
{\it gyrogravitomagnetic ratio} in the same way as common
gyromagnetic ratio, and it should have Dirac value $g=2$
for particle of any spin $J$:
\begin{eqnarray}
\mu_G=J
\label{genc1}
\end{eqnarray}
which coincide with the standard Dirac magnetic moment,
up to the interchange $e \leftrightarrow M$,
making the Bohr magneton equal to $1/2$.

However, the situation changes if one define the gyrogravitomagnetic moment
as a response to the external gravitomagnetic field.
The $\epsilon$ tensor in the coordinate space produces the curl,
and the gravitomagnetic field, acting on the particle spin, is equal to
\begin{eqnarray}
\vec H_J = \frac{1}{2} rot \vec g; \ \vec g_i \equiv g_{0i},
\label{hg}
\end{eqnarray}
where factor $1/2$ is just the mentioned normalization factor in
(\ref{h}). The relevant off-diagonal components of the metric
tensor may be generated by the rotation of
massive gravity source \cite{LL2}.

There is also another effect, induced by this field:
the straightforward analog
of Lorentz force \cite{LL2}, produced by the spin-independent term
in (\ref{e2}). In that case the gravitomagnetic field,
for the low velocity of the particle (such a restriction is actually
inessential, as we can always perform the Lorentz boost, making
the particle velocity small enough) is:
\begin{eqnarray}
\vec H_L = rot \vec g = 2 \vec H_G,
\label{hg1}
\end{eqnarray}

Consider now the motion of the particle in the gravitomagnetic field.
The effect of Lorentz force is reduced, due to the Larmor theorem \cite{Mash},
(which is also valid for small velocity)
to the rotation with the Larmor frequency
\begin{eqnarray}
\omega_L= \frac{H_L}{2}.
\label{hg2}
\end{eqnarray}
This is also the frequency of the {\it macroscopic} gyroscope dragging.
At the same time, the {\it microscopic} particle dragging frequency is
\begin{eqnarray}
\omega_J= \frac{\mu_G}{J}H_J=\frac{H_L}{2}=\omega_L.
\label{hg3}
\end{eqnarray}
The common frequency for microscopic and macroscopic gyroscope
is just the Larmor frequency, so that the gravitomagnetic field
is equivalent to the frame rotation. This should be considered as
a Post-Newtonian manifestation of the equivalence principle.

Let us make here a brief comparison with the literature.
The low energy theorem discussed here
is the necessary ingredient for validity
of gravitational Larmor theorem \cite{Mash}, which otherwise
require an arbitrary assumption about the "classical"
gyrogravitomagnetic
ratio, say, for electron \cite{Mash2}.
At the same time, the equality
of the "classical" and "quantum" frequencies was
found long ago \cite{HD} by comparison of the quantum spin-orbit interaction
with the classical one calculated earlier \cite{Sch}.
Our approach clarify the origin of this equality, as a cancellation
of "geometrical" factor $1/2$ in (\ref{h}) and "quantum"
value $2$ of gyrogravitomagnetic ratio. Note that for free
particle the latter coincides with the usual gyromagnetic ratio, and
such a cancellation provides an interesting connection between
geometry,
equivalence principle and special renormalization properties
(cancellation of
strongest divergencies) for particles with $g=2$ (c.f. \cite{Hripl}).

The crucial factor $1/2$ makes the evolution of the particle helicity
in magnetic and gravitomagnetic fields
rather different. The spin of the (Dirac) particle in the magnetic
field is dragging with the cyclotron frequency, being
twice larger than
Larmor one. It coincides with the frequency of the velocity
precession so that helicity is conserved. At the same time, the
gravitomagnetic field is making the velocity dragging twice
faster than spin, changing the helicity. This factor of $2$, however,
is precisely the one required by the possibility to reduce all the effect
of gravitomagnetic field to the frame rotation. While spin vector
is the same in the rotating frame and is
dragging only due to the rotation of the coordinate axis,
the velocity one is transformed and getting the
additional contribution, providing factor 2 to Coriolis acceleration.

Note that all the consideration is essentially based on the smallness
of the particle velocity, achieved by the mentioned Lorentz boost,
and therefore do not leading to the loss of generality.
There is no doubt about the possibility to construct the
invariant proof, which should result in the Fermi-Walker
transport equation $D_\alpha s^\mu=0$.

Let us consider massive particle scattered by rotating astrophysical
object. The effect of the gravitomagnetic field is reduced
to the rotation of the local comoving frame, which is becoming
inertial at large distances before and after scattering.
Consequently, the helicity is not changed by gravitomagnetic field,
which is confirmed by the explicit calculation of the Born
helicity-flip matrix element in the case of massive neutrino \cite{HD2}.


It may seem, that the equivalence principle should exclude
the possibility of helicity flip in the scattering by gravity source
at all. This is, however, not the case, if usual Newtonian-type
"gravitoelectric"
force is considered \cite{OT99}. Its action is also reduced to the local acceleration
of the comoving frame, in which the helicity of the particle is not altered.
However, the comoving frame after scattering differs from the initial one
by the respective velocity $\delta \vec v = \int \vec a dt$. The
corresponding boost to the original frame
is, generally speaking, changing the helicity
of the massive particle (the similar effect for the gravitomagnetic field
is just the rotation for the solid angle $\delta \vec \Omega =
\int \vec \omega dt$ and does not affect the helicity).
The same boost may be considered as a source of the famous deflection
of particle momentum $\delta \phi \approx |\delta \vec v|/|\vec v|$.
The average helicity of the completely polarized beam after such a
scattering may be estimated in the semiclassical approximation as
$<P> \approx cos \phi \approx 1-\phi^2/2$.
Due to the correspondence principle, this quantity may be expressed as
\begin{eqnarray}
<P>=\frac{d\sigma_{++}-d\sigma_{+-}}{d\sigma_{++}+d\sigma_{+-}} \approx
1-2 \frac{d\sigma_{+-}}{d\sigma_{++}},
\label{P}
\end{eqnarray}
where $d\sigma_{+-} \ll d\sigma_{++}$ - the helicity-flip and
non-flip cross-sections, respectively. Comparing "classical" and
"quantum" expression for $<P>$, one get
\begin{eqnarray}
\frac{d\sigma_{+-}}{d\sigma_{++}} \approx \frac{\phi^2}{4}
\label{+-}
\end{eqnarray}
To check this simple approach, one may perform the calculation
of this ratio for the Dirac particle scattered by the gravitational source.
In the Born approximation, the result is easy to find:
\begin{eqnarray}
\frac{d\sigma_{+-}}{d\sigma_{++}} = \frac{tg^2(\frac{\phi}{2})}{(2\gamma-\gamma^{-1})^2}.
\label{+-e}
\end{eqnarray}


This expression is coinciding with the estimate (\ref{+-}),
as soon as
the deflection angle is small and
the particle is slow
($\gamma=E/m \to 1$), while for the fast particles
\begin{eqnarray}
\frac{d\sigma_{+-}}{d\sigma_{++}} \approx \frac{\phi^2}{16\gamma^2}.
\label{+-r}
\end{eqnarray}

Such an effect should, in particular, lead to the helicity flip
of any massive neutrino. It is very small, when
the scattering by the single object is considered,
but may be enhanced while neutrino is propagating in Universe.
Should the propagation time be large enough, the effect would
result in unpolarized beam of the initially polarized neutrino,
effectively reducing its intensity by the factor of 2.


The manifestation of post-Newtonian equivalence principle
is especially interesting, when "gravitoelectric" component
is absent. Contrary to electromagnetic case, one cannot
realize this situation through cancellation of contributions of
positive and negative charges. At the same time, one may consider
instead the interior of the rotating shell (Lense-Thirring effect).
Especially interesting is the case of the shell constituting
the model of Universe, whose mass and radius are
of the same order, when the dragging frequency may be equal
to the shell rotation
frequency, which is just the Mach's principle \cite{MTW}.
One should note, that the low energy theorem, guaranteeing the
unique precession frequency for all quantum and classical rotators,
is the necessary counterpart of the Mach's principle.

Up to this moment, we considered the gravitational interaction
of the particle, being the eigenstate
of the momentum and spin projection
and described by the
conserved energy-momentum tensor.
Any assumptions
on the particle locality except the locality of energy-momentum
tensor were unnecessary.

We are now ready to postulate the following straightforward generalization \cite{OT99}
of this principle:

{\bf Contributions of quarks and gluons to the
Anomalous Gravitomagnetic Moment of nucleon are zero}

\begin{eqnarray}
\langle P'| T_i^{\mu\nu} |P\rangle=N_i
[2 (P^{\mu} P^{\nu}-\frac{g^{\mu \nu}}{4} M^2) +
\frac{i}{M}P^{(\mu}\epsilon^{ \nu)  \sigma \rho \alpha}
P^\rho S^\sigma \Delta^\alpha]+O(g^{\mu \nu}, \Delta^2).
\label{genc2}
\end{eqnarray}

To test this suggestion let us first turn to perturbative QED analysis. 
The matrix elements of energy momentum tensors of
electrons and photons acquire the logarithmycally divergent contributions,
cancelled in their sum. This problem, at leading order(LO), is similar
to the calculation of QED corrections to gravity coupling \cite{Rob}.
It is sufficient to consider the matrix elements of either electron
or photon energy momentum tensor switched between free electron states,
and the latter case is more simple, being described by the single diagram.
It is enough to consider the terms of zero and first order in $\Delta$.
The divergent contribution to the former is appearing \cite{Rob}
in the traceless part and may be identified 
with the second moment of spin-independent
DGLAP kernel $\int_0^1 dx x P_{Gq}(x)$. The linear term
is known from the orbital angular momentum calculations \cite{Ji,HS,OT98}
and is also equal to that quantity, so that AGM is really zero.


The next important step 
 is to consider the respective finite terms, constituing 
the AGM of electron. It was calculated in QED \cite{Mil,BG},
resulting in the check of the low-energy theorems (\ref{am}). To consider 
the separate AGM of photons and gluons in photon, one should perform
the respective decomposition of 
the total result, and get the non-zero answer, confirmed recently by 
S. Brodsky and collaborators \cite{brod}.    

The general reason for this value was also founded \cite{Mil} as 
emerging due to use the unsubtracted dispersion relations for the 
relevant formfactors, while performing the subtraction, resulting in the 
zero AGM, leads to the absence of the smooth transition $m_e \to 0$ \footnote{I am indebted 
to K. Milton for the illuminating correspondence concerning this issue}. 

Note, however, that these subtraction play the prominent role in the enforcing 
gravitational gauge invariance. While in QED the appearance of gauge invariant 
field strength $F^{\mu \nu}$ corresponds to the subtractions in the relevant amplitudes,
the same should be true for the gravity, where the corresponding gauge-invariant 
quantity is the curvature $R^{\mu \nu}$. As it is quadratic in coordinates(rather than linear, as in 
in QED), the subtraction should make the moments, linear in $\Delta$, equal to zero. 

Performing these subtraction for the separate contributions of various fields while imposing 
Extended Equivalence Principle (EEP) means, that 
such a gauge invariance is postulated for each field  separately. 
The QED example shows, that due to fields interactions the simplest form of such assumption is not valid. 
However, we will now present some arguments in favor of the hypothesis \cite{krim}, 
that in full QCD manifesting such phenomena as confinement 
and chiral symmetry breaking, separate gauge invariance of quarks and gluons (and, therefore, EEP and 
equipartition) may be restored.

Note that due to field interactions the scattering of electron in the external gravitational 
field is accompanied by the emission of soft photons. Let us therefore suggest, that EEP should 
include this emitted photons, and the zero AGM should correspond to the "coherent state" 
of "bare" electron (which is the QED analog of quark contribution to the nucleon AGM) and emitted photon.  
In QCD, due to confinement phenomenon, the emission of soft gluons 
is forbidden and zero AGM of quarks themselves should appear. 
Note that only the singlet combination of quarks could manifest zero AGM, which is 
seen already from the perturbative evolution \ci{OT98}: only the total contribution 
of all quarks, appearing as virtual in QCD beta-function, can preserve equipartition 
at various $Q^2$. 
The equipartition therefore holds at leading order both in QED and QCD.
At NLO order it is violated in QED, and, according to our hypothesis, this corresponds
to the modification of EEP by including the soft emitted photons. In QCD 
equipartition is, in the same line of reasoning, restored due to confinement, 
so that same phenomenon is manifested at LO and beyond the perturbation theory.

The confinement phenomenon is accompanied by chiral symmetry restoration, which,
according to present understanding \cite{for}, is deeply related to confinement and, in particular,
happens at the same temperature in the lattice simulations. This makes the mentioned above 
argument \cite{Mil} against  the subtraction, providing the equipartition, unapplicable to 
full QCD. Indeed, the term $1/m_q$ provided by this subtraction in perturbation theory 
should be replaced by hadronic scale \footnote{More specifically, by the ratio of gluon and quark condensates,
determined by trace anomaly} when chiral symmetry breaking is taken into account.

\section{Higher twist correlations, fragmentation and fracture functions}

Twist 3 quark-gluon correlations (CGC)provide another (and, in QCD, probably the earliest found \cite{ET85})
source of T-odd asymmetries. The transverse gluon field provides both the mass scale and phase shift. 
These effects are in fact quite similar to GPD: in both cases the momenta 
of participating quarks are different: for GPD this difference emerges because of 
diffent hadron mamoenta, while for CGC because of extra gluon. 
This leads, in particular, to the  
similarity between transverse spin in QGC and transverse momentum \cite{PT,Ani1}
of symmetry properties emerging due to T-invariance, found for QGC \cite{ET84}
and GPD \cite{Man}.

While SSA emerged due to high twist correlations involve contributions from both 
large and short distances, their pure non-perturbative sources include T-odd
distribution, fragmentation and fracture functions.  

The most widely known objects are parton distributions, describing
the fragmentation of hadrons to partons and related to the
forward matrix elements
$\sum_X<P|A(0)|X><X|A(x)|P>=<P|A(0)A(x)|P>$
of renormalized non-local light-cone quark and gluon operators.
As they do not contain any variable, providing the cut and
corresponding imaginary phase (to put it in the dramatic manner,
the proton is stable), the T-odd distribution functions can not
appear in the framework of the standard factorization scheme. At
the same time, they may appear effectively, when the imaginary
phase is provided by the cut from the hard process, but may be
formally attributed to the distribution \cite{BMT,ferr}.
Another way of treating the final state interaction, 
found in the explicit model calculations \cite{BHS},
as it was recently stressed by J.C. Collins \cite{col02},
is provided by the path-ordered gluonic exponential.
However, in all cases, the T-odd distribution cannot be universal,
as the imaginary phase appearance depends on the subprocess it is convoluted with. 
To prove the non-univesrality of T-odd distribution $H(x)$, it is sufficient to consider 
its contribution the quark {\it jet} asymmetry in semi-inclusive DIS, where the  
hadronic tensor is proportional to 
\begin{equation}
\label{dist}
W^{\mu \nu} \sim Tr[\gamma^\alpha \gamma^\mu (x \hat P+\hat k_T)\gamma^\nu] H(x) 
\epsilon^{\alpha P S k_T},
\end{equation}
and average it over and $k_T$, assuming that $<k_T^i>=0, <k_T^i k_T^j>=<k_T^2> g_T^{ij}$,
where transverse direction $T$ is defined with respect to the light cone vectors
$P$ and $xP+q$:  
\begin{equation}
\label{disa}
W^{\mu \nu} \sim \frac{<k_T^2>}{Pq} 
H(x) [P^\mu \epsilon^{\nu P S q}+ P^\nu \epsilon^{\mu P S q}].
\end{equation}
The appearance of such a symmetric spin-dependent hadronic tensor is violating 
the real T-invariance. It is interesting that such observable is quite similar
to the one suggested for the search of T-violation and discussed at this conference \cite{barab}.
In both cases one is dealing with correlations of transverse polarization of target nucleon and
tensor polarization of the beam (deuteron in neutron experiment and 
virtual photon in our case) in the total cross-section.  
Note also that (\ref{disa}) is not manifestly electromagnetically gauge invariant,
which is of course true for original expression (\ref{dist}).
Although gauge dependence disappears after contraction with the leptonic tensor,
this signals about necessity to take into account another higher twist effects. 

At the same time, the similar effects for the crossing related
process of semi-inclusive annihilation correspond to the
distributions substituted by fragmentation functions. 
describing
the fragmentation of partons to hadrons and constructed from the
time-like cutvertices of the similar operators
$\sum_X <0|A(0)|P,X><P,X|A(x)|0>$.
As the
latter may contain the imaginary cuts, simulating the
T-violation, the performed calculation is starting to be more
related with physics. Namely, it describes the production of
transverse polarized baryon (one should typically think about
$\Lambda$, whose polarization is easily revealed in its weak
decay) in the annihilation of the unpolarized leptons
\cite{spin96}.

The FRACTURE function (FF)\cite{Tren}, whose particular example
is represented by the diffractive distribution (DD)\cite{coll2},
is related to the object
$\sum_X <P_1|A(0)|P_2,X><P_2,X|A(x)|P_1>,$
combining the properties of FRAgmantation and struCTURE
functions. They describe the correlated fragmentation of hadrons
to partons and vice versa. Originally this term was applied to
describe the quantities integrated over the variable
$t=(P_1-P_2)^2$, while the fixed $t$ case is described by the
so-called extended fracture functions \cite{Gra}.
They may be also extended \cite{TODD}
to describe SSA in such processes.
Namely, such functions can easily get the imaginary phase from the cut
produced by the variable $(P_1+k)^2$. Due to the extra momentum
of produced hadron $P_2$, the number of the possible P-odd
combinations increases. Therefore, they may naturally allow for
the T-odd counterparts. The T-odd fracture function may describe a number of SSA 
at HERMES and, especially, NOMAD \cite{dis02}

\section{Conclusion}

As we see, the T-odd spin asymmetries play a major role in studies of various 
QCD non-perturbative inputs. For GPD, the beam asymmetry in DVCS 
was the main instruments for the experimental check of QCD factorization 
at JLAB and HERMES. 
The studies of SSA at H1 and ZEUS with the polarized lepton may give access 
to twist 3 effects.

The important spin-related aspect of GPD  is represented by their connection to 
the angular momenta of partons. This problem is also related to the post-Newtonian 
Equivalence principle. Its generalization for the separate contributions 
of quarks and gluons which is violated in perturbation theory, but may be restored in full 
QCD due to the confinement and spontaneous chiral symmetry breaking, would lead to the 
equipartition of momentum and angular momentum between quarks and gluons. 

The T-odd distributions may be only effective, as the imaginary phase may be 
associated only with the cut, sepending on the subprocess. The assumption about the
universality of Sivers function leads, after integration over $k_T$ to the real T-violation.
At the same time, the T-odd fragmentation and fracture functions are related to the 
process-indepemdent cuts at large distances and are therefore universal. 

I would like to thank  Organizers for the warm hospitality in Prague.

\end{document}